\documentclass[aps,pra,reprint,showpacs,nofootinbib]{revtex4-1}
\usepackage{amsmath,xcolor,cleveref,graphicx,float,csquotes,amsfonts,epstopdf}
\usepackage[countmax]{subfloat}

\newcommand*{\lbar}[1]{\bar{\bar{#1}}}
\newcommand{\Real}{\textrm{I\!Re}}

\newcommand{\realnumbers}{\textrm{I\!R}}

\DeclareMathOperator{\Tr}{Tr}

\begin{document}
\title{Transformation optics simulation method for stimulated Brillouin scattering}
\author{Roberto Zecca}
\email{roberto.zecca@duke.edu}
\author{Patrick T. Bowen}
\author{David R. Smith}
\author{St\'{e}phane Larouche}
\affiliation{Center for Metamaterials and Integrated Plasmonics and Department of Electrical and Computer Engineering, Duke University, P.O. Box 90291, Durham, North Carolina 27708, USA}
\date{\today}

\begin{abstract}
We develop a novel approach to enable the full-wave simulation of stimulated Brillouin scattering and related phenomena in a frequency-domain, finite-element environment. The method uses transformation optics techniques to implement a time-harmonic coordinate transform that reconciles the different frames of reference used by electromagnetic and mechanical finite-element solvers. We show how this strategy can be successfully applied to bulk and guided systems, comparing the results with the predictions of established theory.
\end{abstract}

\maketitle

\section{Introduction}

In recent years, nonlinear optical phenomena through which light and elastic waves are strongly coupled have garnered a strong interest. Amidst the various phenomena arising from the coupling of optics and elastodynamics, a very well-known and studied example is stimulated Brillouin scattering. Spontaneous Brillouin scattering is a nonlinear optical phenomenon by which light is inelastically scattered by the change in refractive index caused by adiabatic density fluctuations in a medium. These are due to thermal or quantum zero-point effects \cite{Boyd,Fabelinskii}. On the other hand, in stimulated Brillouin scattering (SBS), the density variations are caused by the presence of light, through electrostriction, radiation pressure, and/or optical absorption. It is a third-order optical nonlinearity, whereby elastic and optical waves are coupled in a fluid or solid, mutually exchanging energy \cite{Boyd,Agrawal}. While it has been known and experimented upon for several decades, recent years have seen a renewed interest in SBS and related effects, which already enable many devices, ranging from powerful sources and amplifiers to platforms for the study of slow light \cite{Thevenaz08} and nonreciprocity \cite{Huang11non}, and show great promise for future applications. There has been a growing body of literature on the previously unknown gain enhancements that can be achieved by specifically tailored nanostructures \cite{Rakich12,Qiu13,Dostart15,Wolff15}. In particular, it has been proven by theory and experiment \cite{Shin13} that when waveguide geometries reach the nanoscale, previously unexpected, giant Brillouin gain enhancements occur. These effects are so dramatic as to be 2 to 4 orders of magnitude larger than traditional theories predict (5 in the case of forward SBS) \cite{Rakich12}. Naturally, new theoretical tools have been developed to study and describe these phenomena both in waveguides and various resonant structures \cite{Rakich12,Qiu13,Dostart15}, which are understood to be due to a combination of surface electrostriction and radiation pressure. In hindsight, it is unsurprising that these effects only become predominant at the nanoscale, where the surface-to-volume ratio of particles and waveguides is so high. In these contexts, perturbation theory in the form usually employed in electromagnetism can fail \cite{Johnson02}. These current approaches are all limited to rather simple, or highly symmetric, geometries, such as the aforementioned waveguides and resonators. An alternative to these methods, which require prior knowledge of the modes of the system, are full-wave simulations. One of the most prominent and widely used numeric techniques is the finite-element method. However, as we show, correctly simulating optomechanical effects in the frequency domain is far from straightforward, and the choice of frame of reference for the electromagnetic and mechanical solvers is of critical importance. Na\"{\i}vely overlooking this aspect leads to significantly inaccurate simulation results. In this work, we describe the problem in detail and propose a solution based on transformation optics. The key intuition is that the movement of material points and boundaries can be represented by an effective oscillation of electromagnetic properties. This simulation method is applicable to arbitrarily complex systems and geometries, which can be comprised of several materials, including metals. Thus, it provides an extremely flexible computational platform for the design of optomechanical devices and artificial media, such as plasmonic and metamaterial structures \cite{Eberle08,Smith16}.
	
\section{Vectorial theory of SBS in solids}

In this section, we describe time-harmonic backwards Stokes SBS in a solid medium. Two time-harmonic electromagnetic fields (pump and signal, labeled with the numbers 1 and 2 throughout the paper) counterpropagate in a solid medium, interacting with an elastic wave of angular frequency $\Omega$ and wavevector $\textbf{q}$. For this process, conservation of energy and momentum take the form $\omega_1 = \omega_2 + \Omega$ and $\textbf{k}_1 = \textbf{k}_2 + \textbf{q}$, where $\omega$ and $\textbf{k}$ indicate optical angular frequencies and wavevectors, respectively~\footnote{To extend this treatment to an anti-Stokes process it is sufficient to state energy and momentum conservation as $\omega_1 + \Omega = \omega_2$ and $\textbf{k}_1 + \textbf{q} = \textbf{k}_2$ and to follow the same logical steps.}. For ease of reading, but without loss of generality, we shall assume the solid medium to be isotropic, uniform, and electromagnetically non-dispersive. This last assumption is a reasonable approximation in the case of SBS, where $|\omega_1-\omega_2| \ll \omega_1, \omega_2$. In the following derivations, we consider the behavior of a bi-chromatic, time-harmonic electromagnetic field	
	\begin{subequations}
	\label{eq:defEH}
	\begin{align}
	\tilde{\textbf{E}} &= \tilde{\textbf{E}}_1 + \tilde{\textbf{E}}_2\\
	\tilde{\textbf{H}} &= \tilde{\textbf{H}}_1 + \tilde{\textbf{H}}_2 ,
	\end{align}
	\end{subequations}
where the tilde superscript denotes physical quantities which oscillate rapidly and harmonically in time, and for $n = 1,2$ 
	\begin{subequations}
	\begin{align}
	\tilde{\textbf{E}}_n &= \Real \left(\textbf{E}_n e^{i \omega_n t}\right) = \frac{1}{2} \left(\textbf{E}_n e^{i \omega_n t} + \textbf{E}_n^* e^{-i \omega_n t} \right)\\
	\tilde{\textbf{H}}_n &= \Real \left(\textbf{H}_n e^{i \omega_n t}\right) = \frac{1}{2} \left(\textbf{H}_n e^{i \omega_n t} + \textbf{H}_n^* e^{-i \omega_n t} \right).
	\end{align}
	\end{subequations}
Similarly, the elastic wave is represented by the vectorial displacement and scalar density variation fields
	\begin{subequations}
	\label{eq:defurho}
	\begin{align}
	\tilde{\textbf{u}} =&  \Real \left(\textbf{u} \; e^{i \Omega t}\right) = \frac{1}{2} \left(\textbf{u} \; e^{i \Omega t} + \textbf{u}^*  e^{-i \Omega t} \right) \\
	\Delta \tilde{\rho} =& \Real \left(\Delta \rho \; e^{i \Omega t}\right) = \frac{1}{2} \left( \Delta \rho \; e^{i \Omega t} + \Delta \rho^* e^{-i \Omega t} \right).
	\end{align}
	\end{subequations}
	
	\subsection{Elastodynamics}
	\label{sec:elastodynamics}
	
\begin{figure}
			\centering
			\begin{minipage}[c]{0.48\textwidth}
			\centering
			\includegraphics[width=\textwidth]{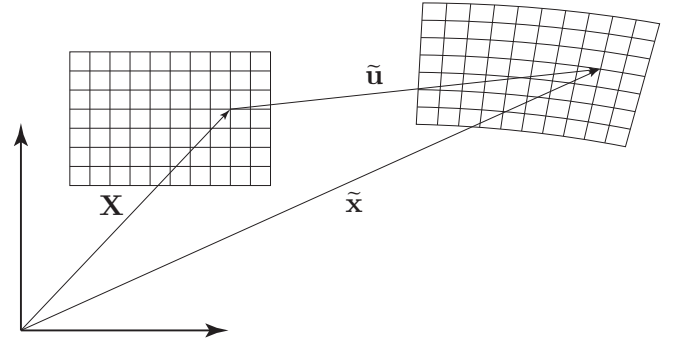}
			\caption{Schematic of the relation between Lagrangian coordinates \textbf{X}, Eulerian coordinates $\tilde{\textbf{x}}$, and displacement $\tilde{\textbf{u}}$.}
			\label{fig:EulerLagrange}
		\end{minipage}
		\end{figure}

Since solids in general support both longitudinal and shear waves, the mechanical aspect of the phenomenon must be described with fully vectorial elastodynamics. For finite-element mechanical simulations, the natural choice for frame of reference are the material (or Lagrangian) coordinates $\textbf{X}$, which index the material points and assign a time-dependent displacement to each, without actually updating their position. By contrast, Eulerian coordinates $\tilde{\textbf{x}}$ follow the position of material points through time. The two frames are related by the displacement $\tilde{\textbf{u}}$ through the relation $\tilde{\textbf{x}} = \textbf{X} + \tilde{\textbf{u}}$, as shown in Fig.~\ref{fig:EulerLagrange}. The Eulerian equilibrium equation, whose form is perhaps more intuitive, is
	\begin{equation}
	\label{eq:equilibriumCauchy}
	\rho \, \frac{\partial^2 \textbf{u}}{\partial t^2} = \nabla_\textbf{x} \cdot \lbar{\sigma} + \textbf{f}_\textrm{v} ,
	\end{equation}
where $\rho$ is the instantaneous mass density, $\lbar{\sigma}$ is the Cauchy stress tensor (referred to the current, deformed, geometry) \cite{Gonzalez}, $\nabla_\textbf{x}$ is the gradient in Eulerian coordinates, and $\textbf{f}_\textrm{v}$ is the sum of body forces (forces per unit deformed volume). Note that the finite linewidth of SBS resonance is due to a mechanical loss term, which can be thought of as included in the definition of stress (by addition of a term proportional to strain rate) \cite{RoyerI}.
In a Lagrangian frame, the equilibrium equations assume instead the form \cite{Gonzalez}:
	\begin{equation}
	\label{eq:equilibriumPK}
	\rho_0 \, \frac{\partial^2 \textbf{u}}{\partial t^2} = \nabla_\textbf{X} \cdot \lbar{P} + \textbf{F}_\textrm{v} ,
	\end{equation}
where $\rho_0$ is the initial mass density, $\nabla_{\textbf{X}}$ is the nabla in Lagrangian coordinates, $\textbf{F}_\textrm{v}$ is the sum of body forces, given with respect to the undeformed volume, and $\lbar{P}$ is the first Piola-Kirchhoff stress tensor (referred to the undeformed geometry). Moreover, $\lbar{P} = \lbar{F} \lbar{S}$, where the $\lbar{F}$ and $\lbar{S}$ tensors are respectively the deformation gradient and the second Piola-Kirchhoff stress \cite{Gonzalez}. It is interesting here to spend a few words on $\lbar{F}$ and its properties. It is defined as the tensor relating Eulerian and Lagrangian coordinates
	\begin{equation}
	\label{eq:definition_F}
	\textrm{d} \textbf{x} = \lbar{F} \, \textrm{d}\textbf{X},
	\end{equation}
and it is a function of the displacement $\tilde{\textbf{u}}$
	\begin{equation}
	\label{eq:Fandu}
	\lbar{F} = \lbar{I} + \nabla_\textbf{X} \tilde{\textbf{u}} ,
	\end{equation}
where $\lbar{I}$ is the identity matrix. The determinant of $\lbar{F}$ is related to the ratio of instantaneous density $\rho$ to the reference (undeformed) density $\rho_0$:
	\begin{equation}
	\label{eq:detF}
	\frac{1}{\det\lbar{F}} = \frac{\tilde{\rho}}{\rho_0} = \frac{\rho_0 + \Delta \tilde{\rho}}{\rho_0} = 1 + \frac{\Delta \tilde{\rho}}{\rho_0} .
	\end{equation}
Traditionally, most nonlinear optics textbooks such as Boyd's \cite{Boyd} model the mechanical aspect of SBS as simple electrostrictive volume forces. However, we want to stress that the computational method we present in this paper can be applied to arbitrarily refined descriptions of optical forces. Following for the moment the traditional description, electrostriction in an isotropic, uniform medium, corresponds to a potential $\tilde{\phi}$ \cite{Boyd}:
	\begin{equation}
	\label{eq:phi}
	\tilde{\phi} = -\frac{1}{2} \epsilon_0 \gamma_\textrm{e} \langle\tilde{\textbf{E}}_1 \cdot \tilde{\textbf{E}}_2\rangle = -\frac{1}{2} \epsilon_0 \gamma_\textrm{e} \, \Real \left( \textbf{E}_1 \cdot \textbf{E}_2^* \, e^{i \Omega t} \right) ,
	\end{equation}	
where the $\langle\cdot\rangle$ sign denotes a time average over an optical period, $\epsilon_0$ is the permittivity of vacuum, and $\gamma_\textrm{e}$ is the electrostrictive constant, defined as \cite{Boyd,Fabelinskii}
	\begin{equation}
	\label{eq:gamma_definition}
	\gamma_\textrm{e} = \left(\rho \, \frac{\partial \epsilon}{\partial \rho}\right)_{\rho = \rho_0} ,
	\end{equation}
where $\epsilon$ is the relative permittivity of the material. The electrostrictive constant relates simply to the photoelastic tensor $\lbar{p}$ in the isotropic case through the fourth power of the refractive index \cite{RoyerI}. The Lagrangian electrostrictive volume force is then
	\begin{equation}
	\label{eq:Lagresf}
	\tilde{\textbf{F}}_\textrm{v} =  -\nabla_\textbf{X} \tilde{\phi} = \frac{1}{2} \epsilon_0 \gamma_\textrm{e} \, \Real \left[ \nabla_\textbf{X}\left( \textbf{E}_1 \cdot \textbf{E}_2^* \right) e^{i \Omega t} \right] .
	\end{equation}
A physical quantity of great importance that must be calculated from the solution to Eq.~\eqref{eq:equilibriumPK} is pressure, which is related to the Cauchy stress tensor through its trace: $p = - \frac{1}{3} \Tr \, \lbar{\sigma}$ \cite{Gonzalez}. Also useful is the variation in mass density $\Delta \tilde{\rho}$, which is related to $\tilde{p}$ through the speed of longitudinal elastic waves $c_p$: $\Delta \tilde{\rho} = \tilde{p} / c_p^2$. The electrostrictive volume force of Eq.~\eqref{eq:Lagresf} can be entered into a finite-element solver as a weak contribution to the elastodynamic partial differential equation, which in frequency domain takes the form
	\begin{equation}
	\label{eq:equilibriumFD}
	- \rho_0 \, \Omega^2 \textbf{u} = \nabla_\textbf{X} \left(\lbar{F} \, \lbar{S}\right) + \frac{1}{2} \epsilon_0 \gamma_\textrm{e} \,  \nabla_{\textbf{X}} \left( \textbf{E}_1 \cdot \textbf{E}_2^* \right) .
	\end{equation}
	
	\subsection{Optics}
	\label{sec:optics}
	
Frequency-domain finite-element electromagnetic solvers are usually cast in Eulerian coordinates, since for most applications there is no need to keep track of mechanical movements at electromagnetic frequencies. For an isotropic, uniform, non-dispersive medium the Eulerian optical wave equation is \cite{Balanis}
	\begin{equation}
	\label{eq:EMTD}
	\nabla^2 \tilde{\textbf{E}} - \frac{n^2}{c^2} \frac{\partial^2 \tilde{\textbf{E}}}{\partial t^2} = \mu_0 \frac{\partial^2 \tilde{\textbf{P}}}{\partial t^2} ,
	\end{equation}
where $n$ is the refractive index, $c$ is the speed of light in vacuum, $\mu_0$ is the permeability of vacuum, and $\tilde{\textbf{P}}$ is a polarization term that acts as a source for the nonlinear process. It can be related to a time-harmonic variation in relative permittivity due to a Brillouin-related scattering mechanism as
	\begin{equation}
	\label{eq:deltaepsilon}
	\tilde{\textbf{P}} = \epsilon_0 \, \Delta \tilde{\chi} \, \tilde{\textbf{E}} = \epsilon_0 \, \Delta \tilde{\epsilon} \, \tilde{\textbf{E}},
	\end{equation}
where $\chi$ is the electric susceptibility of the medium and $\epsilon_0$ is the permettivity of vacuum. In the traditional description of bulk electrostriction, the permittivity variation takes the form $\Delta \tilde{\epsilon} = \gamma_\textrm{e} \, \Delta \tilde{\rho} / \rho_0$ \cite{Boyd}. Representing $\Delta \tilde{\epsilon}$ in the frequency domain and using Eq.~\eqref{eq:defEH} and Eq.~\eqref{eq:defurho}, we can isolate the terms oscillating at $\omega_1$ and $\omega_2$, so that
	\begin{equation}
	\tilde{\textbf{P}} =  \frac{\epsilon_0}{2} \, \Real \left(\Delta \epsilon \, \textbf{E}_2 e^{i \omega_1 t} + \Delta \epsilon^* \textbf{E}_1 e^{i \omega_2 t}\right) .
	\end{equation}
Re-writing Eq.~\eqref{eq:EMTD} in the frequency domain, and separating it into the components oscillating at $\omega_1$ and $\omega_2$, yields:
	\begin{subequations}
	\label{eq:EMFD}
	\begin{align}
	\nabla^2 \textbf{E}_1 + k_1^2 \, \textbf{E}_1 &= -\mu_0 \omega_1^2 \textbf{P}_1 = -\mu_0 \omega_1^2 \frac{\epsilon_0}{2} \Delta \epsilon \, \textbf{E}_2 \\
	\nabla^2 \textbf{E}_2 + k_2^2 \, \textbf{E}_2 &= -\mu_0 \omega_2^2 \textbf{P}_2 = -\mu_0 \omega_2^2 \frac{\epsilon_0}{2} \Delta \epsilon^* \textbf{E}_1 ,
	\end{align}
	\end{subequations}
where we have introduced the scalar wavenumbers $k_1$ and $k_2$, which obey the dispersion relation $k = \omega n / c$, where $n$ is the refractive index of the medium.
						
We have now described the electrostrictive SBS phenomenon through a set of mutually coupled partial differential equations, cast in the form that finite-element software most commonly solve for. However, there is a limitation to simply implementing the nonlinear coupling terms as weak contributions to standard differential equations: a computation solving the electromagnetic wave equation in Eulerian coordinates is not by default able to account for the movement of the geometry, arising from the existence of displacements (which are in turn computed in a Lagrangian frame). This invariably leads to wrong results, especially at the nanoscale, where the effect of (moving) interfaces can play a dominant role \cite{Rakich12,Qiu13,Wolff15}. The obstacle could be avoided by performing a time-domain study instead, but it is in practice undesirable, because of the wildly different time scales of the optical and mechanical periods. Thus, simulating solid-state SBS accurately would seem to be unreasonably onerous from a computational standpoint in the time domain and outright impossible in the frequency domain.
		
\section{Transformation optics as a route to frequency-domain SBS simulations}

\subsection{Transformation optics for a moving frame}
To circumvent the difficulty described in the previous section, a possible strategy is to employ transformation optics (TO) \cite{Ward96,Pendry06,Leonhardt06} in a way that enables a standard electromagnetic solver to correctly account for the moving frame. The idea is to represent the movement of material points and boundaries by an effective time-oscillation of electromagnetic properties. In TO, the material properties of an original (isotropic, to simplify the discussion) medium, unprimed in Eq.~\eqref{eq:TO}, are transformed through the following relation:
	\begin{subequations}
	\label{eq:TO}
	\begin{align}
	\lbar{\epsilon} \, ' =& \frac{\lbar{A} \lbar{A}^T}{\det \lbar{A}} \, \epsilon = \lbar{g} \, \epsilon \\
	\lbar{\mu} \, ' =& \frac{\lbar{A} \lbar{A}^T}{\det \lbar{A}} \, \mu = \lbar{g} \, \mu ,
	\end{align}
	\end{subequations}
where $\lbar{A}$ is the Jacobian matrix of the coordinate transformation and $\lbar{g}$ is the metric tensor in three dimensions. The transformed permittivity $\lbar{\epsilon}$ and permeability $\lbar{\mu}$ are in general complex, symmetric rank-two tensors. In our case, the transformation is between a moving frame (Eulerian) and a fixed frame (Lagrangian). The Jacobian of this transformation is the deformation gradient $\lbar{F}$ mentioned in Sec. \ref{sec:elastodynamics}. Using Eq.~\eqref{eq:defurho}, Eq.~\eqref{eq:Fandu}, and Eq.~\eqref{eq:detF}, it is possible to obtain 
	\begin{equation}
	\lbar{F} = \lbar{I} + \Real \left[\left(\nabla \textbf{u}\right) e^{i \Omega t} \right] ,
	\end{equation}
and thus a compact expression for the metric tensor $\lbar{g}$
	\begin{equation}
	\label{eq:metric}
	\lbar{g} = \sum\limits_{n = -3}^{3} \lbar{g}_n e^{i n \Omega t} ,
	\end{equation}
with the property $\lbar{g}_n = \lbar{g}_{-n}^*$, reflecting the fact that, as expected, the metric maps real coordinates to real coordinates. More details on the derivation and explicit expressions for the metric components can be found in Appendix \ref{sec:appendix}.

\subsection{Wave-like equations for non-dispersive materials}

\begin{figure*}
			\centering
			\includegraphics[width=\textwidth]{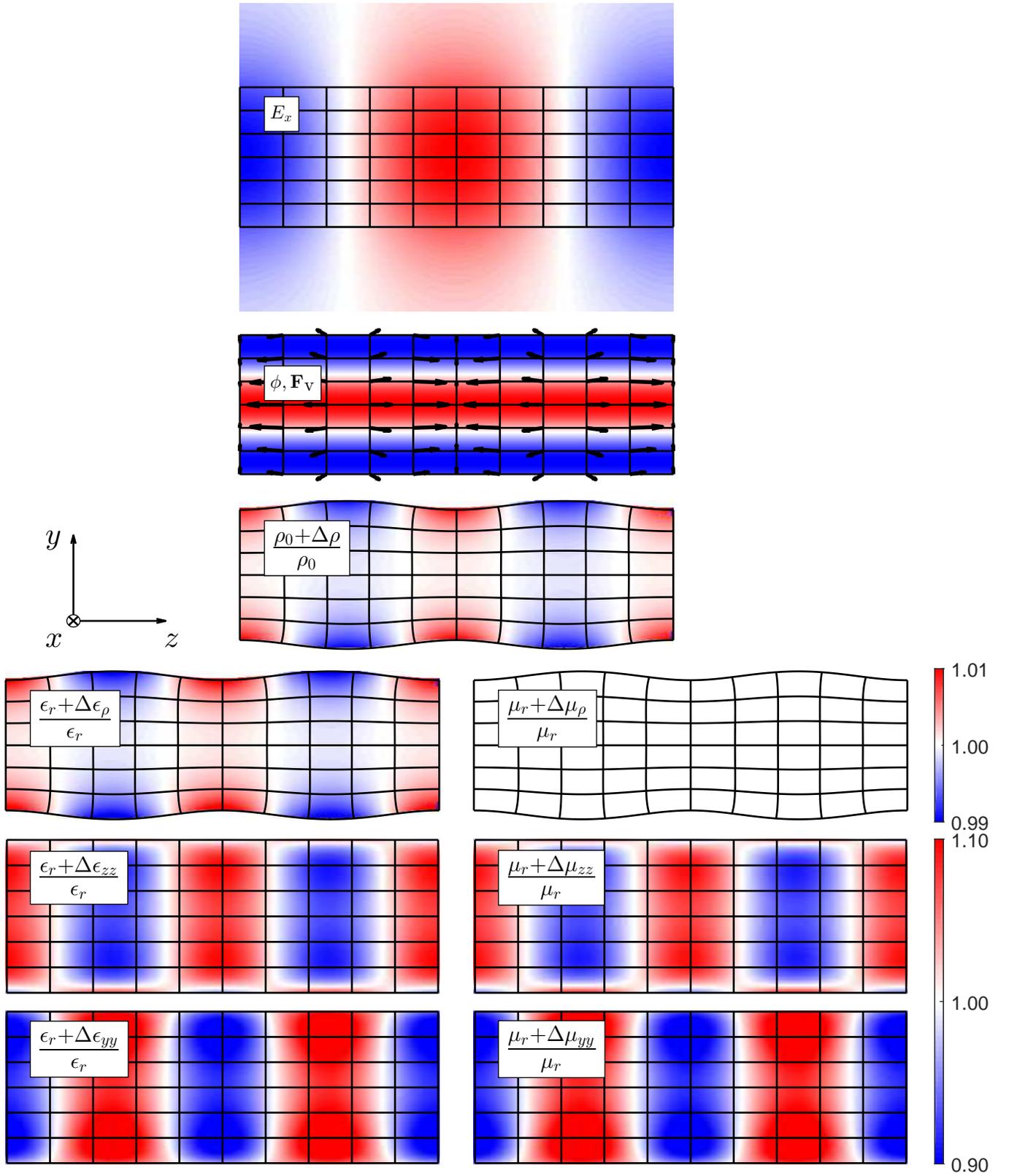}
			\caption{(Color online) Conceptual schematic of TO method, as applied to the system described in Sec. \ref{sec:2D}. Two optical TE$^z$ guided modes (one of which is depicted in the 1$^{\textrm{st}}$ row) counterpropagate in a dielectric slab waveguide, giving rise to a mechanical potential $\phi$ and the corresponding force field $\textbf{F}_\textrm{v}$ (2$^{\textrm{nd}}$ row, color map and arrows, respectively). The force excites one or more elastic modes (3$^{\textrm{rd}}$ row, warped grid), thus creating a mass density variation field $\Delta \rho$ (3$^{\textrm{rd}}$ row, color map). This in turn induces a relative permittivity variation field $\Delta \epsilon_\rho$ (4$^{\textrm{th}}$ row, left column), but there is no effect $\Delta \mu_\rho$ on permeability in the case of ordinary diamagnetic optical materials (4$^{\textrm{th}}$ row, right column). With our method, we calculate effective anisotropic properties (5$^{\textrm{th}}$ and 6$^{\textrm{th}}$ rows, left column permittivity, right column permeability) that enable the simulation of SBS coupling while keeping material points fixed (Lagrangian frame). All figures are depicted in reference to a given time $t_0$.}
			\label{fig:BIG}
		\end{figure*}

For anisotropic, inhomogeneous material properties such as the ones typically yielded by transformation optics, it is not possible to obtain an equation in a form as simple as a Helmholtz wave equation. For time-independent properties one can derive an equation that resembles Helmholtz's, some form of which is in fact the master equation in many full-wave frequency-domain finite-element solvers
	\begin{equation}
	\label{eq:COMSOLmasterRF}
	\nabla \times \left(\lbar{\mu}_r^{-1} \nabla \times \textbf{E}\right) - k_0^2 \left(\lbar{\epsilon}_r - \frac{i \lbar{\sigma}_\textrm{e}}{\omega \epsilon_0}\right) \textbf{E} = \textbf{0} ,
	\end{equation}
where $\lbar{\sigma_\textrm{e}}$, $\lbar{\epsilon}_r$, and $\lbar{\mu}_r$ are respectively the electrical conductivity, relative permittivity, and relative permeability tensor. Our goal is to obtain a similar result in the case of time-dependent transformed material properties, in the form shown in Eq.~\eqref{eq:TO} and Eq.~\eqref{eq:metric}. The fundamental ideas of our method, however, are not necessarily tied to this form. In fact, they are in principle applicable to any frequency-domain finite-element formulation of electromagnetism. A conceptual schematic of the method is presented in Fig.~\ref{fig:BIG}.
Let us first consider the well-known differential, macroscopic form of the charge-free Maxwell's equations:
	\begin{subequations}
	\label{eq:Maxwell}
	\begin{align}
	\label{eq:Faraday}
	-\nabla \times \tilde{\textbf{E}} &= \frac{\partial \tilde{\textbf{B}}}{\partial t}\\
	\label{eq:Ampere}
	\nabla \times \tilde{\textbf{H}} &= \frac{\partial \tilde{\textbf{D}}}{\partial t} + \tilde{\textbf{J}}_\textrm{e} \\
	\nabla \cdot \tilde{\textbf{D}} &= 0 \\
	\nabla \cdot \tilde{\textbf{B}} &= 0 ,
	\end{align}
	\end{subequations}
where $\tilde{\textbf{E}}$ is the electric field, $\tilde{\textbf{D}}$ is the electric flux field, $\tilde{\textbf{H}}$ is the magnetic field, $\tilde{\textbf{B}}$ is the magnetic flux field, and $\tilde{\textbf{J}}_\textrm{e}$ is the electric current density. Supposing the undeformed medium is isotropic we can rewrite Eq.~\eqref{eq:Faraday} and Eq.~\eqref{eq:Ampere} by applying the appropriate TO rules \cite{Ward96,Kundtz09}:
	\begin{subequations}
	\begin{align}
	\label{eq:FaradaySBS}
	- \nabla \times \tilde{\textbf{E}} &= \frac{\partial}{\partial t} \left(\lbar{g} \mu \tilde{\textbf{H}}\right)\\
	\label{eq:AmpereSBS}
	\nabla \times \tilde{\textbf{H}} &= \frac{\partial}{\partial t} \left[\lbar{g} \left( \epsilon + \epsilon_0 \Delta \epsilon \right)\tilde{\textbf{E}}\right] + \lbar{g} \sigma_\textrm{e} \tilde{\textbf{E}} ,
	\end{align}
	\end{subequations}
where $\mu, \epsilon, \sigma_\textrm{e} \in \realnumbers$.  In fact, for traditional materials at optical frequencies, in first approximation $\mu = \mu_0$. Traditional models of SBS model $ \Delta \epsilon = \gamma_\textrm{e}\Delta \tilde{\rho}/\rho_0 $ (cf. Sec. \ref{sec:optics}), but we will not specify a form for $\Delta \epsilon$, which can act as a ``black box'' for any relevant scattering mechanism involved. Using Eq.~\eqref{eq:defEH}, a few simple algebraic steps allow us to calculate $\lbar{g} \mu\textbf{H}$, admitting the Stokes conservation of energy and that we can disregard all terms not oscillating at $\omega_1$ or $\omega_2$:
	\begin{align}
	&\lbar{g} \mu \tilde{\textbf{H}} = \\*
	\nonumber
	&\mu \Real \left[e^{i \omega_1 t} \left(\lbar{g}_0 \textbf{H}_1 + \lbar{g}_1 \textbf{H}_2\right) + e^{i \omega_2 t} \left(\lbar{g}_1^* \textbf{H}_1 + \lbar{g}_0 \textbf{H}_2\right)\right] .
	\end{align}
It is thus possible to rewrite Eq.~\eqref{eq:FaradaySBS}, separating the terms at different frequencies and switching to a more compact frequency-domain notation:
	\begin{subequations}
	\label{eq:newFaraday}
	\begin{align}
	- \nabla \times \textbf{E}_1 &= i \omega_1 \left(\lbar{A} \, \textbf{H}_1 + \lbar{B} \, \textbf{H}_2\right) \\
	- \nabla \times \textbf{E}_2 &= i \omega_2 \left(\lbar{B}^* \textbf{H}_1 + \lbar{A} \, \textbf{H}_2\right) ,
	\end{align}
	\end{subequations}
where $\lbar{A} = \lbar{g}_0 \mu$ and $\lbar{B} = \lbar{g}_1 \mu$. Since $\mu \in \realnumbers$ as discussed earlier, $A_{ij} \in \realnumbers$ too. We can then left-multiply Eq.~\eqref{eq:newFaraday} by $\lbar{A}^{-1}$ to get
	\begin{subequations}
	\label{eq:invA}
	\begin{align}
	- \lbar{A}^{-1} \nabla \times \textbf{E}_1 &= i \omega_1 \left(\textbf{H}_1 + \lbar{A}^{-1} \lbar{B} \, \textbf{H}_2\right)\\
	- \lbar{A}^{-1} \nabla \times \textbf{E}_2 &= i \omega_2 \left(\lbar{A}^{-1} \lbar{B}^*  \textbf{H}_1 + \textbf{H}_2 \right) .
	\end{align}
	\end{subequations}
We can now take the curl of each side of Eq.~\eqref{eq:invA} to get, invoking the linearity of the curl operator,
		\begin{subequations}
			\label{eq:curlcurl}
			\begin{align}
			- \nabla \times \lbar{A}^{-1} \nabla \times \textbf{E}_1 &= i \omega_1 \left[\nabla \times \textbf{H}_1 + \nabla \times \left(\lbar{A}^{-1} \lbar{B} \, \textbf{H}_2\right) \right]\\*
			- \nabla \times \lbar{A}^{-1} \nabla \times \textbf{E}_2 &= i \omega_2 \left[\nabla \times \textbf{H}_2 + \nabla \times \left(\lbar{A}^{-1} \lbar{B}^* \textbf{H}_1\right) \right] .
			\end{align}
		\end{subequations}
Analogously to the derivation of Eq.~\eqref{eq:newFaraday}, we can obtain expressions for the curl of the magnetic field complex amplitudes, taking into account the linear and nonlinear polarization terms, and the conduction current:
	\begin{subequations}
	\label{eq:newAmpere}
	\begin{align}
	\nabla \times \textbf{H}_1 &= i \omega_1 \left[ \left( \lbar{C} + \lbar{K} \right) \textbf{E}_1 + \left( \lbar{D} + \lbar{L} \right) \textbf{E}_2 \right] \\*
	\nabla \times \textbf{H}_2 &= i \omega_2 \left[ \left( \lbar{D}^* + \lbar{L}^* \right) \textbf{E}_1 + \left( \lbar{C} + \lbar{K} \right) \textbf{E}_2 \right] ,
	\end{align}
	\end{subequations}
where $\lbar{C} = \lbar{g}_0 \, \epsilon$, $\lbar{D} = \lbar{g}_1 \epsilon$, $\lbar{K} = \epsilon_0 \left(\lbar{g}_1 \Delta\epsilon^* + \lbar{g}_1^* \Delta\epsilon \right)/2$, and $\lbar{L} = \epsilon_0 \left( \lbar{g}_0 \Delta\epsilon + \lbar{g}_2 \Delta\epsilon^* \right)/2$. We have folded conductivity into permittivity as is customary, making the latter complex $\epsilon = \epsilon' - i \epsilon''$ with $\epsilon''>0$ for optically lossy materials. Moreover, $K_{ij} \in \realnumbers$. Substitution of Eq.~\eqref{eq:newAmpere} into Eq.~\eqref{eq:curlcurl} yields Eq. \ref{eq:wavelike}.
\begin{widetext}
\begin{subequations}
\label{eq:wavelike}
\begin{align}
\nabla \times \lbar{A}^{-1} \nabla \times \textbf{E}_1  - \omega_1^2 \lbar{C} \textbf{E}_1 &= \omega_1^2 \left[\lbar{K} \textbf{E}_1 + \left(\lbar{D}+ \lbar{L}\right) \textbf{E}_2 \right] - i \omega_1 \nabla \times \left(\lbar{A}^{-1} \lbar{B} \textbf{H}_2 \right) \\*
\nabla \times \lbar{A}^{-1} \nabla \times \textbf{E}_2  - \omega_2^2 \lbar{C} \textbf{E}_2 &= \omega_2^2 \left[ \left(\lbar{D}^* + \lbar{L}^* \right) \textbf{E}_1 + \lbar{K} \textbf{E}_2 \right] - i \omega_2 \nabla \times \left(\lbar{A}^{-1} \lbar{B}^* \textbf{H}_1 \right)  .
\end{align}
\end{subequations}
\end{widetext}
Upon comparison with Eq.~\eqref{eq:COMSOLmasterRF}, we notice that the general form of the equations is preserved. Naturally, Eq.~\eqref{eq:wavelike} are mutually coupled through the right-hand sides as a consequence of the nonlinear process they describe. On the left-hand sides, $\lbar{A}$ takes the place of $\lbar{\mu}_r$, and $\omega_n^2 \lbar{C}$ that of $k_0^2 \left[\lbar{\epsilon}_r - i \lbar{\sigma}_\textrm{e}/\left(\omega \epsilon_0\right)\right]$.

\section{Examples of applications}
\label{sec:applications}

In this section, we present two applications of our method to predict SBS gain in well-understood solid-state systems. In the first example, we consider a one-dimensional (1-D) amplifier setup. In the second case, we highlight how more refined descriptions of optical forces can be incorporated into the method, allowing it to accurately predict gain enhancement in nanostructures, as described in \cite{Rakich12,Qiu13,Wolff15}. All simulations are run in \textsc{comsol} Multiphysics 5.2, with the full-wave electromagnetic solver master equation replaced by Eq.~\eqref{eq:wavelike}, except where noted.

	\subsection{1-D solid-state SBS amplifier}

\begin{figure}
					\centering
					\begin{minipage}[c]{0.48\textwidth}
					\centering
					\includegraphics[trim=0cm 0.25cm 0cm 0.25cm, clip=true, width=\textwidth]{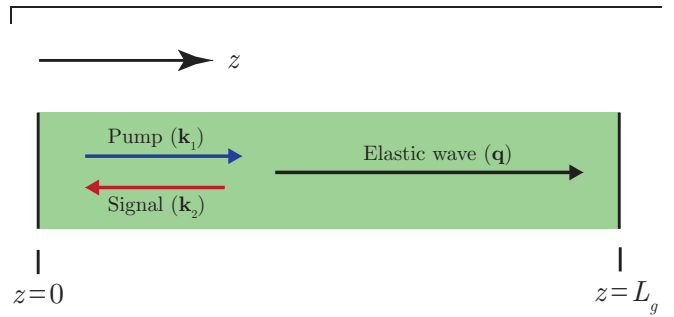}
					\caption{(Color online) 1-D backward SBS amplifier: schematic of 2-D simulation (inspired by \cite{Boyd}). The lateral boundaries are connected through periodic boundary conditions, making the domain effectively infinite in the transverse direction. Open boundary conditions generate optical fields at one end ($z=0$ for the pump, $z=L_g$ for the signal) and transmit them without reflection at the other. Elastic waves are generated by optical forces, and absorbed at either $z-$boundary by perfectly matched layers. $L_g$ is the characteristic gain length. Elastic waves are computed in the plane strain approximation.}
					\label{fig:1Dampsetup}
			\end{minipage}
			\end{figure}

\begin{figure}
			\includegraphics{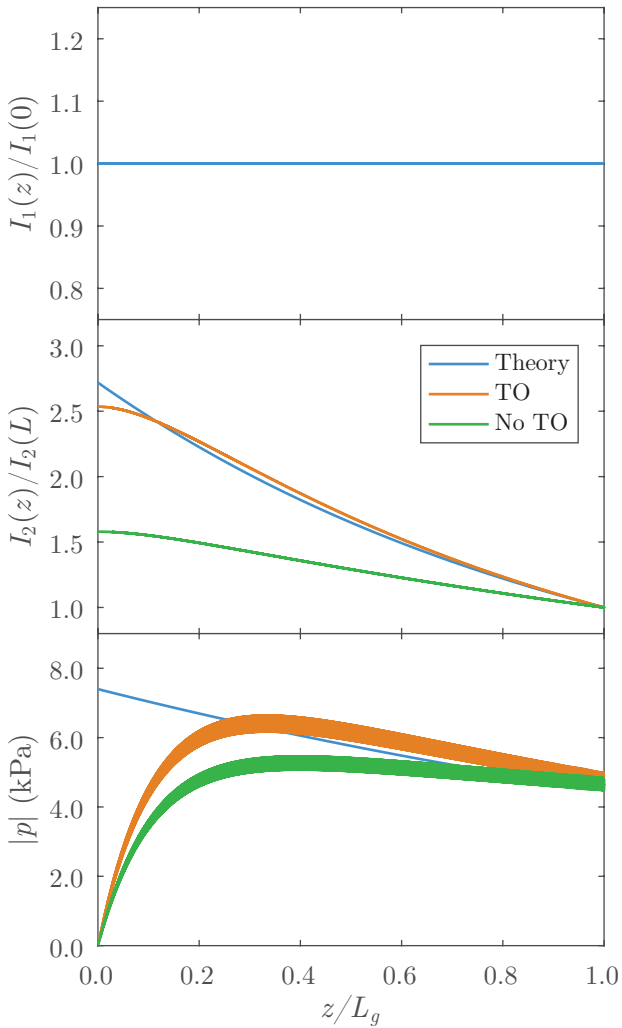}
			\caption{(Color online) 1-D backward SBS amplifier at resonance: (top) relative pump intensity (middle) relative signal intensity, as predicted by theory (blue), the TO method (orange), and a simply coupled simulation (green) (bottom) pressure amplitude.}
			\label{fig:1Dampresult1}
		\end{figure}

\begin{figure}
			\centering
			\includegraphics{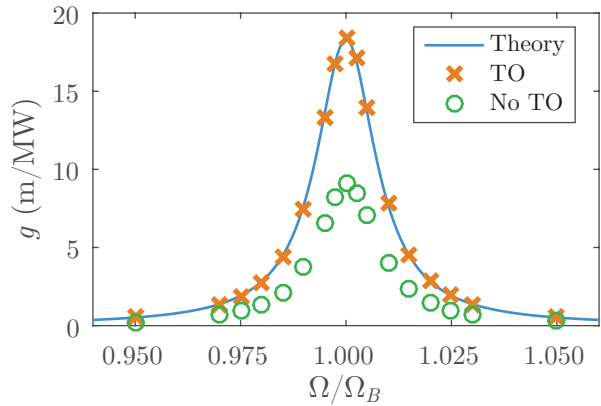}
			\caption{(Color online) 1-D backward SBS amplifier: exponential gain $g$ spectrum predictions: theory (continuous blue line) against simulations run with TO method (orange ``x'' series) and without (green ``o'' series).}
			\label{fig:1Dampresult2}
		\end{figure}

	As a preliminary demonstration of the effectiveness of the method, we simulate a simple solid-state 1-D backward SBS amplifier. It consists of two counter-propagating electromagnetic waves in a solid Brillouin medium that is finite in the propagation direction $\hat{z}$, and infinite in the other two. A pressure wave arises due to optical forces, in this case the standard bulk electrostriction mentioned in Sec. \ref{sec:elastodynamics}. The first wave, the pump, is chosen to be much more intense than the signal seed ($I_1 \gg I_2$), so we can expect the undepleted pump approximation to be valid. In this case, signal amplification is described appropriately by a simple exponential model, i.e. the solution to the ordinary differential equation $\tfrac{\partial}{\partial z}  I_2 \left(z\right) = -g \, I_1 \, I_2 \left(z\right)$ \cite{Boyd}. The simulation is described in further detail in Fig.~\ref{fig:1Dampsetup}.
	Simulations were run over a range of mechanical frequencies $\Omega$, keeping the pump frequency $\omega_1$ constant and adapting the signal frequency as $\omega_2 = \omega_1-\Omega$. Results from the simulation at the resonant frequency $\Omega = \Omega_B$ are displayed in Fig.~\ref{fig:1Dampresult1}. The top panel shows that the pump intensity remains constant throughout the propagation distance, thereby confirming the validity of the undepleted pump approximation. In the middle panel, results for relative signal intensity are reported for simulations run with and without the TO method, and are compared with theory \cite{Wolff15}. The graph highlights how simply implementing the nonlinear coupling into the software is inadequate, and how our method is necessary to obtain a solution consistent with theory. The difference between theory and simulation with the TO method in the left-hand side of the graph is easily interpreted as a transient feature: the electrostrictive force only acts within the simulation region, i.e. over a finite length. Therefore, the pressure wave must build up gradually, as shown in the bottom panel of Fig.~\ref{fig:1Dampresult1}, before taking the trend predicted by theory (that instead concerns itself with plane waves, which exist and are coupled over the whole propagation space). Since pressure mediates the energy transfer from pump to signal, this explains small the deviation in $I_2$.
	For each simulation, a relative signal intensity graph such as the one in Fig.~\ref{fig:1Dampresult1} is generated. The data is then fitted with an exponential function $I_2 \left(z\right) = I_2 \left(L\right) \textrm{exp}\left[I_1 g \left(L-z\right)\right]$ \cite{Boyd}, from which the gain factor $g$ is extracted. These values are plotted in Fig.~\ref{fig:1Dampresult2} against the theoretical prediction. The agreement between the two approaches is excellent, whereas the simply coupled simulations fail to predict the scale of the Lorentzian resonance peak.
		
	\subsection{Dielectric elastic slab waveguide at different scales}
	\label{sec:2D}
		
\begin{figure}
			\centering
			\begin{minipage}[c]{3.4in}
			\includegraphics[trim=0cm 0cm 0cm 0cm, clip=true, width=\textwidth]{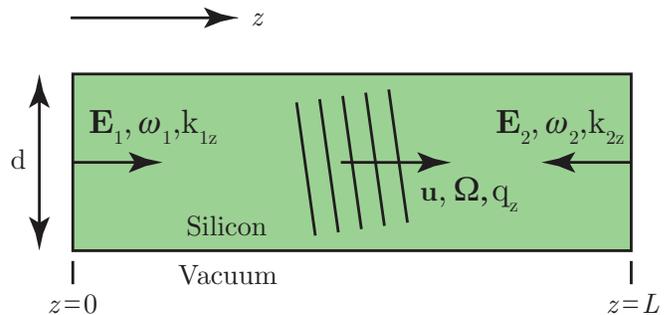}
			\end{minipage}
			\caption{(Color online) Dielectric slab SBS amplifier: schematic of the waveguide, of finite thickness $d$. The direction of propagation is $z$, and the problem is translationally invariant in the out-of-plane direction, making it 2-D.}
			\label{fig:2Dampsetup}
		\end{figure}

\begin{figure}
			\centering
			\includegraphics{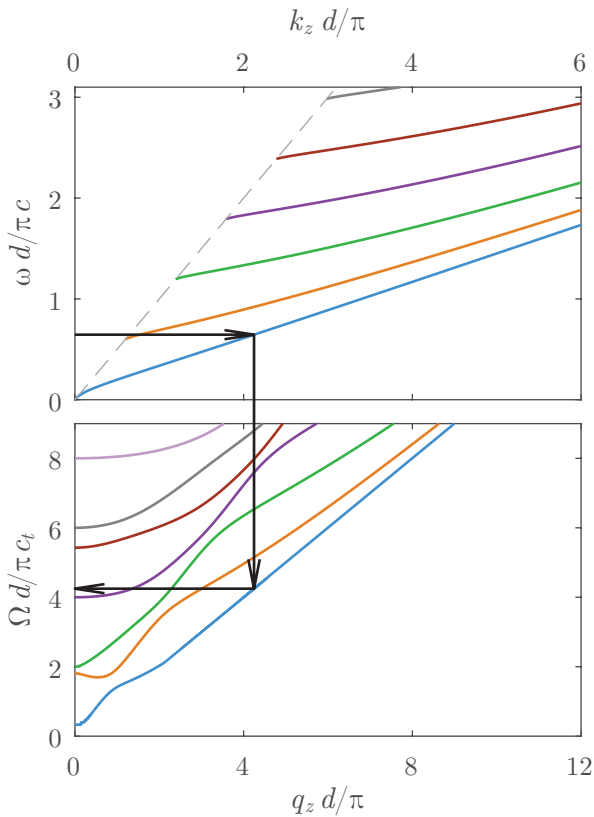}
			\caption{(Color online) Dielectric slab SBS amplifier: schematic of the dispersion diagrams for lossless TE$^{z}$ electromagnetic and longitudinal elastic waves. In simulations, a fixed optical frequency $\omega$ is picked. Selecting a waveguide thickness $d$, one can read off the corresponding propagation wavenumber $k_z$ for the desired mode (the lowest-order in our case). By phase-matching, the corresponding elastic propagation constant $q_z \simeq 2 k_z$ is determined, from which one finds the frequency $\Omega$ of the desired elastic mode.}
			\label{fig:dispersion}
		\end{figure}

\begin{figure}
			\centering
			\includegraphics{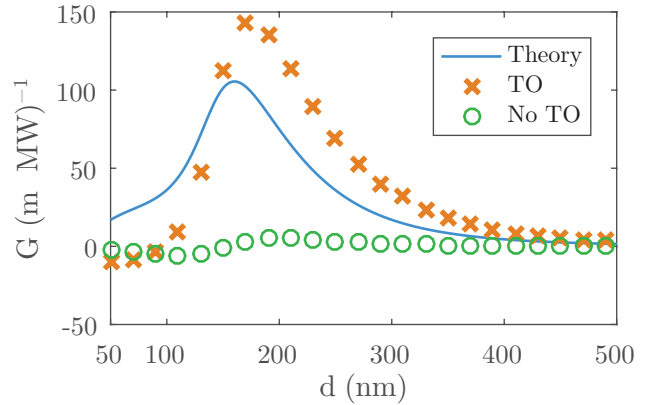}
			\caption{(Color online) Dielectric slab SBS amplifier: gain $G$ versus waveguide thickness $d$. Theory (continuous blue line); simulations with TO method (orange ``x'' series) and without method (green ``o'' series).}
			\label{fig:2Dresults}
		\end{figure}

	We next apply our method to a structured system: a suspended slab waveguide of finite thickness and infinite extent in the plane, as depicted in Fig.~\ref{fig:2Dampsetup}. The example is conveniently simple, because it possesses translational invariance in the plane perpendicularly to the direction of propagation, thus making the problem effectively 2-D. We study the backward SBS interaction between the fundamental TE mode and the quasi-longitudinal elastic modes, all of which share a plane of symmetry at half thickness (with respect to the electric field and longitudinal displacement).
	The dispersion diagrams were computed semi-analytically from the waveguide dispersion relations \cite{Balanis,Achenbach} and are depicted in Fig.~\ref{fig:dispersion}. For a broad range of waveguide thickness values, we simulate SBS at the optical free-space wavelength of 1.55 $\mu$m, in an undepleted-pump regime, with the previously presented selection rule applied to propagation constants $k_{1z} = k_{2z} + q_z $ and operating at the elastic frequency of the lowest-order elastic mode.
	The waveguide material is silicon, whose properties are modeled as follows: relative permittivity $\epsilon_r=12.25$, relative permeability $\mu_r=1$, photoelastic coefficient $p_{21}=0.017$, Young's modulus $E_Y=170$ GPa, Poisson's ratio $\nu=0.28$, mass density $\rho_0=2329 \, \textrm{kg m}^{-3}$. The material is assumed to be optically lossless, while all elastic modes are arbitrarily assigned an isotropic loss factor of $1/200$, which translates into a viscosity tensor whose nonzero elements are $1/200$ of the corresponding stiffness tensor elements. From the simulations we extract a combined gain value with a procedure similar to the one outlined in the previous section. An important \textit{caveat} is that, in this case, we adopt a definition of gain more suited to guided systems, i.e.  $\tfrac{\partial}{\partial z}  P_1 \left(z\right) = \tfrac{\partial}{\partial z}  P_2 \left(z\right) = -G \, P_1 \left(z\right) \, P_2 \left(z\right)$, where $P_{1,2} \left(z\right)$ are, respectively, the time-averaged guided pump and signal powers. Provided $P_1 \gg P_2 \; \forall z$, $P_1$ can be treated as a constant (undepleted pump approximation). Optical forces due to electrostriction (both as a volume force on the bulk and as a pressure term on the boundaries) and radiation pressure are taken into account \cite{Rakich12,Qiu13}. In Fig.~\ref{fig:2Dresults} we compare the results of simulations, run with and without the TO method, with those of the most advanced theory of SBS available in the literature, that of Wolff \textit{et al} \cite{Wolff15}.
	The simulation results have qualitatively similar trends, although the TO method predicts a peak value of gain that is larger by approximately one order of magnitude. This highlights the importance of taking into account the movement of material points and boundaries when performing SBS calculations. The agreement between TO method and the Wolff theory, however, is much closer. The discrepancies can be attributed to the Wolff theory being strictly monomodal with respect to each field, whereas the TO-SBS simulations by their nature take into account all elastic modes at the chosen frequency, be they propagating or evanescent, that contribute constructively (destructively) to the SBS process, thereby increasing gain (losses). In this fashion, our method is able to predict configurations in which the combined contribution of elastic modes does not give rise to any gain, but instead result in net loss. Thus, our method expands the prediction capabilities of current theories, which by construction are only able to predict positive values of gain in optically lossless systems \cite{Rakich12,Qiu13,Wolff15,Wolff15power}.
	
\section{Conclusions}

	We describe a TO-based strategy to enable the finite-element simulation of SBS phenomena in the frequency domain. The method is versatile in that it can work with any kind of geometry, and can incorporate arbitrarily refined descriptions of optical forces. Furthermore, it does not require prior analytic or modal knowledge of the problem. The method is readily generalizable to anisotropic background materials and to a fully tensorial description of the photoelastic effect. Future developments may include extending applicability to fluid domains, which are usually described by either a scalar pressure field, or more generally by a vectorial velocity field. Our method provides a powerful platform for the design of artificial media, in particular metamaterials and plasmonic systems, whose electromagnetic and elastic properties (including resonances) can be engineered with ample control.
	
\begin{acknowledgments}
\noindent The authors would like to thank Professors Daniel J. Gauthier and Robert L. Bryant for valuable discussions on nonlinear optics and differential geometry, respectively. This work was financially supported by the W911NF-09-1-0539 Transformation Optics MURI grant.
\end{acknowledgments}

\appendix
\section{Metric tensor components}
\label{sec:appendix}

By substituting the expressions for $\lbar{F}$ as a function of displacement Eq.~\eqref{eq:Fandu} and $\det \lbar{F}$ as a function of density variation Eq.~\eqref{eq:detF}
	\begin{subequations}
	\begin{align}
	\lbar{F} &= \lbar{I} + \Real \left[\left(\nabla_\textbf{X} \textbf{u}\right) e^{i \Omega t} \right] \\*
	\det \lbar{F} &= \left[1 + \Real \left(\Delta \rho \; e^{i \Omega t} /\rho_0 \right)\right]^{-1}
	\end{align}
	\end{subequations}
into the TO formula $\lbar{g} = \lbar{F} \lbar{F}^T / \det \lbar{F}$ , one obtains $\lbar{g} = \sum_{n = -3}^{3} \lbar{g}_n e^{i n \Omega t} $ by writing
	\begin{align}
	\lbar{g} =& \left\{\lbar{I} + \Real \left[\left(\nabla \textbf{u}\right) e^{i \Omega t} \right] \right\} \left\{\lbar{I} + \Real \left[\left(\nabla \textbf{u}\right)^T e^{i \Omega t} \right] \right\} \times \\*
	\nonumber
	& \times \left[1 + \Real \left(\Delta \rho \; e^{i \Omega t} /\rho_0 \right)\right] ,
	\end{align}
where the \textbf{X} subscript has been dropped from the gradient for ease of reading. In particular, the four metric coefficients are
\begin{widetext}
\begin{subequations}
	\begin{align}
	\lbar{g}_0 &= \lbar{I} + \frac{1}{2} \Real \left\{\left(\nabla \textbf{u}\right) \left(\nabla \textbf{u}\right)^\dagger + \frac{\Delta \rho}{\rho_0} \left[\left(\nabla \textbf{u}\right)^* + \left(\nabla \textbf{u}\right)^\dagger \right] \right\}  \\
	\lbar{g}_1 &= \frac{\left(\nabla \textbf{u}\right) + \left(\nabla \textbf{u}\right)^T}{2} + \frac{\Delta \rho}{2 \rho_0} \left\{\lbar{I} + \frac{1}{2} \Real \left[\left(\nabla \textbf{u}\right)\left(\nabla \textbf{u}\right)^\dagger \right] \right\} + \frac{\Delta \rho^*}{\rho_0} \frac{\left(\nabla \textbf{u}\right) \left(\nabla \textbf{u}\right)^T}{8} \\
	\lbar{g}_2 &= \frac{\left(\nabla \textbf{u}\right) \left(\nabla \textbf{u}\right)^T}{4} + \frac{\Delta \rho}{\rho_0} \frac{\left(\nabla \textbf{u}\right) + \left(\nabla \textbf{u}\right)^T}{4} \\
	\lbar{g}_3 &= \frac{\Delta \rho}{\rho_0} \frac{\left(\nabla \textbf{u}\right) \left(\nabla \textbf{u}\right)^T}{8} ,
	\end{align}
	\end{subequations}
where the $\dagger$ sign indicates the conjugate transpose operator.
\end{widetext}

\end{document}